# CHANGE MANAGEMENT AND VERSION CONTROL OF SCIENTIFIC APPLICATIONS

Bojana Koteska and Anastas Mishev

Faculty of Computer Science and Engineering, Ss. Cyril and Methodius University, Skopje, Macedonia
bojana.koteska@finki.ukim.mk
anastas.mishev@finki.ukim.mk

*ABSTRACT*

*The development process of scientific applications is largely dependent on scientific progress and the experimental research results. Thus, dealing with frequent changes is one of the main problems faced by the developers of scientific software. Taking into account the results of the survey conducted among scientists in the HP-SEE project, the implementation of change management and version control software processes is inevitable. In this paper, we propose software engineering principles that should be included in the development process to improve the version control and change management. Moreover, we give some specific recommendations for their implementation, thereby making a slight modification of already generally accepted templates and methods. The development steps practiced by scientists should not be replaced completely, but they need to be supplemented with appropriate practices, documents and formal methods. We also emphasize the reasons for the inclusion of these two processes and the consequences that may arise as a result of their non-application.*

## 1. INTRODUCTION

The purpose of the scientific application development is to solve a particular problem in any scientific field and it is usually intended for use in the research community or by the scientist himself. Commonly, the application developer is a scientist who has no formal training for learning programming techniques and skills, but he learns them independently [1]. Taking into consideration the aforementioned facts and the results of a survey we conducted among scientists in HP-SEE project and the results obtained from the research presented in [1] and [2], it can be concluded that current scientific application development practices do not include the software engineering development principles. There are three different user communities in this project: Computational Physics (software developers from 6 SEE countries contributed with 8 applications), Computational Chemistry (software developers from 6 SEE countries contributed with 7 applications, collaborating with scientists from more than 20 research centres in Europe) and Life Sciences(software developers from 5 SEE countries contributed with 7 applications) [3].

Scientific applications are characterized by frequent changes which occur as a result of misunderstandings between software engineers and scientists [4] and research advancements. Thus, for the purpose of dealing with changes, the inclusion of version control and change management processes as officially accepted practices for software development is essential.

In order to change and improve the current scientific development practices, the goal of this paper is to suggest some specific software engineering practices that will help the implementation of version control and change management processes. We also propose a change management plan template and some version control realization ideas. The objectives of this paper are to emphasize the need and to show the benefits of including software engineering principles and software tools for version control and change management of the development process. These recommendations will be helpful for the following reasons: to reduce the number of software bugs; to deal with possible changes; to have a more detailed view of the software evolution process which is important; especially when new members will join the team; to provide better organization of the development activities and to increase overall quality of the scientific applications.

The motivation for writing this paper comes from the results obtained from the survey conducted among the scientists that are part of the project [5]. The results showed that the majority of the development teams do not perform software release version tracking and only few development teams think that it is an important process. Software engineering practices and quality standards can improve the development process of scientific applications by: providing more organized development activities and better management process that lead to reducing the development time; increasing the application quality; reducing the number of errors; etc.

The paper is organized as follows: The background work is presented in the second section. The third section describes the specific development process of scientific applications. In the fourth section, the reasons for performing version control and change management are presented and also the consequences of their non-application. In the next section we propose some specific software engineering practices for improving the software version control and change management plan. We also give some reasons why they cannot be used in the same form in the scientific software development process. In the sixth section, we present a template for change management plan and some software tools for performing the version control. In the last section, we give conclusion and describe our future work.

## 2. BACKGROUND

Several authors have been writing about version control and change management for scientific software development, but they did not propose any specific practices for their implementation or templates. In [6], the author presented the results from an interview about version control of scientific software and according to the answers there are interviewees who do not use version control, some of them use it without any specific tools and some of them do it consistently with using appropriate tools. Other examples of lack of knowledge by scientists about version control systems are presented in [7].

In [8], the authors stated the benefits from change management and using version control systems. They mentioned: reducing loss of information, raising the actuality of non-code artifacts and monitoring the impact of changes.

In [9], an overview of several change management models that are primarily intended for development of traditional commercial software is given. The authors also proposed a spiral model which is organized in 4 cycles: specification of a new requirement request or a problem with existing products, problem solving (non-technical viewpoint), system engineering and technology-specific (technical solution).

In [10], the authors show that the old way of sending codes by e-mails or by using Dropbox, should be replaced with the use of software version control systems. The authors point out systems for performing software version control. These systems provide an option for multiple users to develop all the source code in parallel and to synchronize it later.

The results obtained from the survey [5] confirmed the non-use of formal software engineering practices and as a consequence a number of development teams were not sure whether their software contains bugs or not. The bad development practices are also confirmed by the results from two other surveys [11] [2].

In [12], the authors mention the change management as an important process when bug tracking systems are used in coding and testing phases of the software development. Change management and tracking changes are useful also for the risk management process [13].

## 3. SCIENTIFIC APPLICATIONS DEVELOPMENT

One of the definitions of scientific applications, describes the scientific application as a software that performs a simulation of real-world activities by using mathematical models and formulas [14]. This definition indicates that scientific applications are not intended for commercial use, but for the advancement of science and scientific research.

Since the scientific application is not intended for commercial use and no other users except scientists exist, the process of defining software requirements is not easy. On the other hand, software engineers often have problems when they want to know the requirements at the beginning of the software development process and scientists do not know them and do not want to write them in a formal form [15]. The requirements definition can be written using specific words in the scientific domain, which are not known for software engineers [16] and usually the additional explanation of the terms is necessary.

The problem of a concise and timely requirements definition imposes a variety of new problems in the later stages of the development process, such as: incorrect test definition or no test definition, delayed testing process, improper error handling, lack of proper implementation plan, occurrence of frequent changes, creation of new software versions which depends on the new requirements definition, etc.

The different cultures and understandings of software engineers and scientists [15] are one of the biggest problems for software development that follows the principles and practices of software engineering.

A conflict of interests occurs when scientists believe that the most important part is to achieve scientific progress and software engineers to develop software correctly. If software works right, scientists do not care if the code is optimized or any possible modifications should be made. Scientists are primarily interested in solving complex problems and mathematical calculations and they do not care how the optimization is performed unless there is free memory space. Checking whether the software works is different from the standard testing process because the validation is performed usually by comparing the results obtained from physical experiments [2].

Not dealing with frequent changes which usually happens because of the scientific progress can result in errors. If version software control is not done properly, only one mistake can destroy the whole software application. This is especially important in the development process of systems that can endanger a human life or nature.

Non-application of release version tracking is also confirmed by the results from the survey among HP-SEE participants and it is shown in Fig.1.

The above facts indicate that there is a big difference in the scientific and commercial application development process. Incorrect software development and non-inclusion of software engineering principles have a negative impact on application quality. The development process of scientific software should be flexible, but software engineering practices should be included where possible.

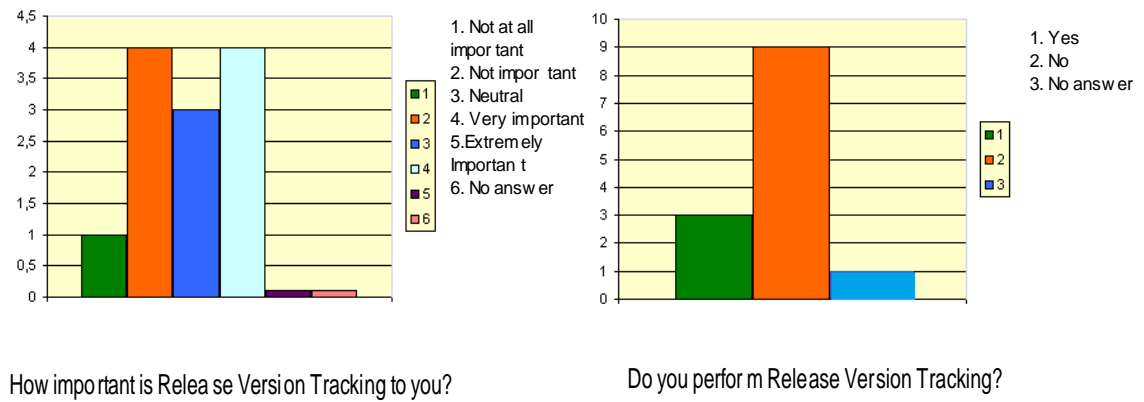

Figure 1. HP-SEE survey results - Release version tracking

## 4. WHY USE CHANGE MANAGEMENT AND SOFTWARE VERSION CONTROL?

Software change management supports software changes and altering of the software after changes by providing software operation and maintenance [17]. This process is especially important in the final software development stage and its primary goal is to ensure the consistency of the system abstraction levels and parts [18]. One of the reasons for performing the software change management are the constant code changes and changes in requirements that can cause any further changes in software [19].

A change management plan contains information about the change management process and it is a part of the project management plan [20]. It provides a central information system about software changes which is its main benefit [9].

It is useful to create a preliminary change management plan based on previous experience and information about the change types that have occurred before in the software development aimed at solving a similar problem. The main benefit of the change management plan is possibility software to be returned to a previous stable version when something goes wrong.

Software version control is a process of identifying and keeping track of different software versions and releases. It provides management of the source code, electronic documents, paper documents, executable code, bitmap graphics, and other artifacts [21]. The responsibility of the version control system is the resolution of the conflicts which occur as a result of concurrent changes to the same code sections. This system must provide consistent coupling of the changed parts [22]. It is also useful when some code changes and improvements should be made (source code optimization, statement reordering and parallelization).

## 5. SOFTWARE ENGINEERING PRACTICES FOR CHANGE MANAGEMENT AND VERSION CONTROL OF SCIENTIFIC APPLICATIONS

Given the fact that scientists usually do not apply software engineering practices in the scientific application development process [15][5], they do not consider software version control and change management as important factors for developing quality applications. There are some already accepted software engineering practices which are mostly used in the commercial application development process, but they are still not used in the scientific software development. These practices are primarily oriented to use of specific standards, templates, methods, etc. The research showed that scientists are not familiar with change management and version control process and tools.

The process of making changes in the development process of scientific software needs to be customized according to the current practices of scientific software development, but software engineering principles should be included also. This approach is important so that scientists can easier accept a change and more successful cooperation between the software engineer and scientist can be established.

One of the key questions that must be answered before some changes are being implemented in the development process of scientific applications is how to adapt to software engineering change management practices and software version control. The first thing that needs to be done is to determine the characteristics of scientific software and to identify differences in the development process between the commercial and scientific applications. Next, the existing practices for change management and software version control should be modified in order to cover all the specifics of scientific software. In addition, the basic development concepts that are already accepted as standards of quality should be followed. Modifications must enable flexible change management process and software version control. Version control software should to be amended in accordance with the application needs.

The model cycle presented in [9] is shown in Fig.2.

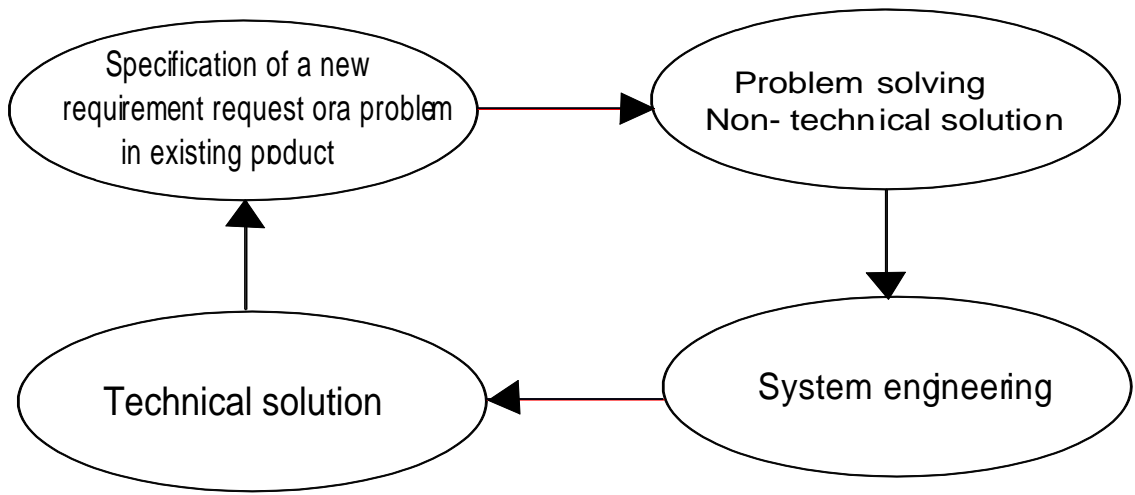

Figure 2. Change management model

This model can be easily adapted to change management in scientific software development. Several minor modifications should be made, i.e. some parts of the 4th cycles should be removed and some steps in each of the cycles should be added. We propose the following modifications:

- The steps: "define an interest group" should be removed (no special interest group exists);

- The step "acceptance testing" should be replaced with "validating the results using the results obtained from the physical experiment" or if it is not possible with "validating the results using theory or universally accepted solutions"; (no real users exist);

- A new step "coordination between scientists and software engineers" should be added at the end of the 2nd cycle.

The process of scientific software version control should be done by using software tool that provide a user-friendly, collaborative and efficient environment. One of the most important things to consider when choosing the version control tool is the ability to provide coordination between changes made on the same code sections at a time and visual representation of the changes made in each version compared to the previous.

Non-application of the software engineering practices in all development phases, especially change management process that occurs as a result of code error correction or software upgrading leads to the creation of low quality applications and unsolved problems. This kind of approach must be replaced by the inclusion of software engineers and professional testers in the development team and changing the bad habits. They will show the scientists how to use some methods, formal engineering principles and tools to automate some of the developing activities. Since no real users exist, the acceptance testing should be replaced by validation using the results obtained from experiments. Also, the coordination between scientists and software engineers is an important factors for the successful development process.

## 6. CHANGE MANAGEMENT AND VERSION CONTROL IN PRACTICE

### 6.1. Change Management Plan for Scientific Applications

Change management plan is an important step that needs to be done before the process of application development starts. In scientific software development the need for creating such a plan is indispensable because changes are an integral part of the application development and they usually occur in unexpected time. One of the goals of this paper is to propose change management plan for scientific application development.

We propose the following structure of the change management plan document:

**Project description**. This is the section where the problem that need to be solved is explained and also some specifics about its development from logical and technical points of view are briefly described.

**Glossary**. In this part the terms used in the whole document should be defined. The words used in the scientific domain should be included also. The Glossary is required mostly for software engineers that will join the development team in the future.

**Change control responsibilities**. The responsibilities of the development team members that are part of the change management process should be specified. In addition, each member must get specific tasks for different types of changes such as: persons responsible for change identification, change implementation, risk identification, writing code status, etc.

**Change management approach**. This section describes the process of dealing with changes, i.e. an approach that should be used to handle each change. The changes should be divided according to their type. If there are more different types, a certain solution for each change type should be proposed. The process of dealing with changes should be described in detail, starting from a change request to their realization. This section must be supplemented constantly when any change is approved and implemented. This information can be used as official documentation and that will be useful when new a member joins the development team. The document should contain change description, change scope, change impact, details about its implementation, possible risks and state of the code before and after the change for each change (software version control can be used for getting the data).

**Change description**. The description of the change shall be given in this section. The problem that has to be solved by this change should be described in more detail.

**Change scope**. The reason for making the current change should be described in more detail and the benefits of that change should be stated.

**Change impact**. The change impact must be stated in order to express all the consequences that arise as a result of the code changing.

**Software status before/after the change implementation**. This part should be used to show the differences between the code segments that are affected by the current change. If a software version control system is being used, then it can help the process of showing the code status before and after the change.

**Change implementation**. It explains the technical solution of the change problem. This step is especially useful for the software engineers and scientists that will join the development process later.

**Risks identification**. This part is intended to identify all the risks that can appear as a result of change implementation. Each of the possible risks should be described in details. The following information should be written: risk type, affected code segments, modules or implementation logic and future modifications that might happen.

**Change validation**. Software validation is the process of testing. This section should shortly describe the testing method used to check the functionality of the system after the change is being implemented and the final results should also be presented. If the physical experiment is performed, then the conclusion of the comparison between the experimental results and final application results should be given.

The change management plan must be created before the software development starts and it must be gradually upgraded. A good strategy of dealing with changes allows much more control of the development activities from the beginning and it provides higher quality of the change management process and the whole scientific application. In addition, the developers acquire good developing habits and they adopt some basic software engineering practices which are part of the software quality standards.

## 6.2. Version Control Tools Selection

No universally accepted version control tool for scientific application development exists. In order to select the most appropriate software tool for version control, the development team must define the needs and goals of the scientific application. It is important to state all the constraints because that will help the selection process. The experience of the scientific community members for developing applications of similar type can help when choosing a tool.

The tool selection depends on different factors such as: application size, number of team members, the need for details, log and report types, visual representation of changes, etc.

Some of the open source available tools for version control designed for different operating system environments:

- Git - distributed version control tool available for multiple platforms: Linux, Windows, Mac OS X, Solaris. It provides cheap local branching, convenient staging areas, and multiple workflows [23].

- Subversion - central version control system which is part of the Apache Software Foundation and it provides reliable, simple, and safe environment [24].

- CVS - client-server version control system for Unix-like and Windows operating systems. It provides source and document changes recording and flexible modules database [25].

CVS is an older version control system and it is rarely used today. For users who prefer faster and distributed solution we recommend Git. Otherwise, users who want more centralized access and organization, and easier maintenance process should use the Subversion system.

## 7. CONCLUSIONS

In this paper, we have outlined the software engineering practices that should be used for change management and version control of scientific applications. Inspired by the results of the survey which showed that scientists do not care much about the process of dealing with changes, we also proposed a change management plan template. Our research has shown that change management and version control are crucial steps in the development process, especially in a development environment when changes happen often. The proposed template and modification of the existing change management model can be easily adapted to any scientific application development process. Using software engineering practices will increase the quality standards in the scientific application development.

Our future research will be oriented to propose a framework for scientific application development, where more detailed and specific software engineering practices for different types of applications will be given for each of the development phases. This framework should provide a concept for developing quality scientific applications.

## ACKNOWLEDGEMENTS

This work was supported in part by the European Commission under EU FP7 project HP-SEE (under contract number 261499).

## REFERENCES

[1]     J. Segal, (2008) "Models of scientific software development", in SECSE 08, First International Workshop on Software Engineering in Computational Science and Engineering (Leipzig, Germany).

[2]     V.R. Basili. et al, (2008) "Understanding The High Performance Computing Community: A Software Engineer's Perspective", *IEEE Software*, Vol. 25, pp29-36.

[3]     HP SEE official web site, http://www.hp-see.eu/

[4]     J. Segal, (2003) "When software engineers met research scientists: a field study", technical report, Department of Computing, Faculty of Mathematics and Computing, The Open University, Walton Hall, Milton Keynes, United Kingdom.

[5]     B. Koteska & A. Mishev, (2013) "Software engineering practices and principles to increase quality of scientific applications", in: Markovski, S., Gusev, M. (Ed.), *ICT Innovations 2012, Advances in Intelligent Systems and Computing*, Springer Berlin Heidelberg, Vol. 207, pp245-254.

[6]     R. Sanders, (2008) "The development and use of scientific software", MSc thesis, School of Computing, Queens University, Kingston, Ontario, Canada.

[7]     G. Wilson, (2006) "Where's the real bottleneck in scientific computing?", *American Scientist*, Vol. 94, pp5-6.

[8]     V. Hoffmann, H. Lichter & A. Nyen, (2009) "Processes and practices for quality scientific software projects", *Proceedings of 3rd International Workshop on Academic Software Development Tools WASDeTT-3*, Antwerp, pp95-108.

[9]     S. M. Ghosh, H. R. Sharma & V. Mohabay, (2011) "Analysis and modeling of change management process model", *International Journal of Software Engineering and Its Applications*, Vol. 2, pp123-134.

[10]    G. Wilson, et al, (2012) "Best practices for scientific computing", CoRR, abs/1210.0530.

[11]    J. E. Hannay, et al, (2009) "How do scientists develop and use scientific software?", *in Proceedings of the 2009 ICSEWorkshop on Software Engineering for Computational Science and Engineering*, IEEE Computer Society Washington, DC, USA, pp1-8.

[12]    R. Kumar, et al., (2013) "Improving software quality assurance using bug tracking system", *International Journal of Computer Science and Information Technologies*, Vol. 4, pp492-497

[13]    M. Elmaallam, A. Kriouile, (2011) "Towards a model of maturity for is risk management", *International Journal of Computer Science & Information Technologies*, Vol. 3, pp171-188

[14]    Definition of scientific application, http://www.pcmag.com/encyclopedia term /0,,t=&i=50872,00.asp, Accessed: 10 March, 2013.

[15]    J. Segal, (2008) "Scientists and software engineers: A tale of two cultures", *in Proceedings of the Psychology of Programming Interest Group*, University of Lancaster, UK, pp 44-51.

[16]    D. Sloan, et al., (2009) "User research in a scientific software development project", *in Proceedings of the 23rd British HCI Group Annual Conference on People and Computers: Celebrating People and Technology*, British Computer Society, pp423-429.

[17]    P. Habela & K. Subieta, (2002) "OODBMS metamodel supporting configuration management of large applications", *in Proceedings of the 8th International Conference on Object-Oriented Information Systems*, pp40-52.

[18]    M. Makarainen, (2000) "Software change management process in the development of embedded software", academic diss., technical research centre of Finland, Finland.

[19]    C. Letondal & U. Zdun, (2003) "Anticipating scientific software evolution as a combined technological and design approach", *in Second International Workshop on Unanticipated Software Evolution (USE2003)*, Warsaw, Poland, pp1-15.

[20]    C.S. Snyder (2013) *A project manager's book of forms: a companion to the PMBOK,* Guid.John Wiley & Sons, United States.

[21]    B.B. Agarwal, S. P. Tayal (2009) *Software engineering and testing*, Jones & Bartlett Learning, USA.

[22]    T. Keller, Open source version control, BSc thesis, Department Computer Science, Mathematics & Natural Sciences, University Of Applied Sciences, Leipzig, 2006.

[23]    Git official website, http://git-scm.com/

[24]    Subversion official website, http://subversion.apache.org/

[25]    CVS official website, http://cvs.nongnu.org/